\documentclass[aps,pra,showpacs,twocolumn,superscriptaddress]{revtex4}

\usepackage{graphicx}
\usepackage{dcolumn}
\usepackage{bm}
\usepackage{amsfonts}
\usepackage{amsmath}
\usepackage[usenames]{color}
\usepackage[]{float}
\usepackage{subfloat}
\usepackage{subfig}
\usepackage{caption}
\newcommand{\bra}[1]{\left\langle{#1}\right\vert}
\newcommand{\ket}[1]{\left\vert{#1}\right\rangle}
\newcommand{\beq}{\begin{equation}}
\newcommand{\eeq}{\end{equation}}
\newcommand{\bea}[1]{\begin{equation}\begin{array}{#1}}
\newcommand{\eea}{\end{array}\end{equation}}
\newcommand{\beqn}{\begin{eqnarray}}
\newcommand{\eeqn}{\end{eqnarray}}

\begin{document}

\title{Quantum correlations as precursors of entanglement}

\author{A. Auyuanet}
\email{auyuanet@if.ufrj.br}

\author{L. Davidovich} 
\affiliation{Instituto de F\'{\i}sica, Universidade Federal do Rio de
Janeiro, Caixa Postal 68528, Rio de Janeiro, RJ 21941-972, Brazil}

\begin{abstract}
We show that for two initially excited qubits, interacting via dipole forces and with a common reservoir, entanglement is preceded by the emergence of quantum and classical correlations.  After a time lag, entanglement finally starts building up, giving rise to a peculiar entangled state, with very small classical correlations.  Different measures of quantum correlations are discussed, and their dynamics are compared and shown to lead to coincident values of these quantifiers for several ranges of time.
\end{abstract}


\maketitle

\section{INTRODUCTION}
\label{sec:introduction}

The characterization of entanglement and the elucidation of its role in quantum computation remain formidable challenges, in spite of the  conspicuous presence of this concept in quantum physics since the fundamental and instigating papers published by Einstein, Podolski, and Rosen, as well as Schr\"odinger, in 1935 \cite{epr35,schrodinger35}.  
Motivation for the understanding of this subtle concept is stimulated not only by its fundamental character, but also by its applications in quantum information~\cite{nielsen00} and, much specially, by the perspective that entanglement could be the key ingredient in the  increased efficiency of quantum computing compared to classical computation for certain quantum algorithms \cite{nielsen00,ekertjozsa,jozsalinden}.

In an entangled state, classical and quantum correlations may coexist. Indeed, for instance, in the singlet state $|\Psi_-\rangle=(|10\rangle-|01\rangle)/\sqrt{2}$ there is a perfect classical correlation between the first and the second qubit, namely if the first is in state $|1\rangle$ the second is in state $|0\rangle$, and vice-versa. 
Several measures of entanglement have been proposed, like the concurrence \cite{wooters98} and the negativity \cite{ziczkowski98}.  Also, different criteria for the existence of classical correlations have been proposed \cite{Henderson02,vedral08}.
In Ref.~\cite{vedral08}, the existence of genuine classical correlations was associated to the non-vanishing of $n$-party correlation functions involving local observables of the system. Based on this definition, it  was shown that it is possible to have multiparty entangled states with no genuine classical correlations, as long as the number of parties is larger than two \cite{vedral08}.

More recently, it has become clear that entanglement does not exhaust the realm of quantum correlations. Indeed, separable states can exhibit quantum correlations, which seem to play a role in the explanation of the power of some schemes of quantum computation \cite{knill98,poweronequbit,Lanyon08}. 
Several quantifiers have been proposed for these quantum correlations, starting with the work by Ollivier and Zurek \cite{olliverzurek}, who introduced the ``quantum discord". Intuitively, this measure quantifies, in a bipartite system, the minimum change in the state of the system and on  the information of one of its parts induced by a measurement on the other part.  For a state with zero quantum discord, it is possible to measure any of its parts without changing the state of the system.  An example would be the state
\begin{equation}\label{classicalcor}
\rho_{1}=\frac{1}{2}\left( \ket{0}\bra{0}\otimes \ket{+}\bra{+}+\ket{1}\bra{1}\otimes \ket{-}\bra{-} \right)\,,
\end{equation}
for which there exist one-dimensional complete projective measurements on both the first and second subsystem -- namely, measurements on the basis $\{|0\rangle,|1\rangle\}$ for the first qubit and $\{|+\rangle,|-\rangle\}$ for the second qubit --  that do not perturb the overall quantum state, nor the state of each part. The quantum discord of this state is zero: the state has only classical correlations.

On the other hand, for the state
\begin{equation}\label{quantumcor}
\rho_{2}=\frac{1}{2}\left( \ket{0}\bra{0}\otimes \ket{+}\bra{+}+\ket{1}\bra{1}\otimes \ket{0}\bra{0} \right)\,,
\end{equation}
it is clear that any projective measurement performed on the second qubit disturbs the state of the first qubit. In this case, quantum discord is
different from zero. The same is true for the maximally-entangled singlet state, which is transformed into a statistical mixture upon measurement of any of the two qubits.

An interesting implication of this concept is the demonstration that vanishing quantum discord is necessary and sufficient for completely positive maps~\cite{shabani}, that is, it is not necessary to assume, in the usual master equation description, that the initial state of system+environment is a product state. It could be a more general separable state, as long as the quantum discord vanishes.
Quantum discord has been calculated for several families of quantum states and compared with the entanglement \cite{aliX}. Modified versions have been proposed, with different physical meanings \cite{discorddemons}.

Other measures of quantum correlations, involving non-orthogonal measurements (positive operator-valued measurements - POVM's) and measurements on both qubits, have been introduced in the literature \cite{HendVedral,luomid,Molmer}, and have been studied and compared with quantum discord in different situations: the DQC1 model of mixed-state quantum computation \cite{dattamixed}, and within the context of accelerated frames \cite{dattaaccelerated}. We present in Sec. \ref{sec:Quantum and Classical Correlations} these other quantifiers, as well as a precise definition of quantum discord. 
The existence of quantum correlations in the absence of entanglement is another subtle trait of quantum mechanics, still to be fully understood. 

In this paper, we show that when two qubits interact with each other through a common environment and also through dipole forces, there is a peculiar dynamical behavior intertwining quantum correlations, classical correlations, and entanglement. For  two qubits initially excited, quantum correlations show up as precursors of entanglement, growing up and then shrinking as entanglement belatedly appears. At this moment, classical correlations, even though non-vanishing, become practically negligible. The dynamics of quantum correlations is studied in terms of two distinct quantifiers, the above-mentioned quantum discord and the measured-induced disturbance (MID), introduced by Luo \cite{luomid}. In particular, we show that the MID coincides with the  quantum discord for part of the evolution.

The dynamics of multiparty entangled systems interacting with independent environments that act on each of its parts is quite different from the time-dependent behavior of each individual component of the system. Quite generally, the decay of entanglement is non-exponential, and it may lead to separability at finite times, before each part reaches its final state~\cite{simon02,dood04,yu1,carvalho04,mintert05b,fine05,yonac061,santos06,yu06,yu062, almeida07,eberlyscience}. The finite-time disappearance of entanglement, sometimes called ``sudden-death of entanglement''~\cite{yonac061,eberlyscience} was experimentally demonstrated by Almeida et al.~\cite{almeida07}. 
The dynamics of quantum correlations other then entanglement has also been analyzed, both theoretically~\cite{maziero,mazzola} and experimentally~\cite{xu}.

As the two qubits get closer, at a distance comparable to the radiated wavelenght,  a different physical situation arises. The model of individual and independent environments  does not correspond anymore to the physical reality: one must consider that the qubits interact with the same environment, and furthermore the interaction between the qubits, which depends on their physical characterization, must also be taken into account. This has important consequences for the dynamical behavior of the system.
A common environment may entangle initially separable systems, even in the absence of direct interaction
\cite{braun,kimheat,jakobczyk,schneider,luodecoprod,davies,derkacz,jakojamspont,benatti,ficekspontaneous2,ficekdark}. This can be easily understood from the fact that the product state $|0\rangle|\otimes|1\rangle$ may be expressed as a sum of a singlet and a triplet component:
\begin{equation}
|01\rangle={1\over2}[(|01\rangle-|10\rangle)+|01\rangle+|10\rangle)]\,,
\end{equation}
and, since the singlet component does not decay under a common environment, a residual entanglement remains at asymptotic times~\cite{malena}.

This is not valid anymore for two initially excited atoms, since then the singlet component does not appear. In this case, one can show that the system remains separable throughout its evolution~\cite{malena}.
The situation changes dramatically, however, when dipole interactions between the two initially excited atoms are taken into account. In this case, it was shown that entanglement does emerge, but only after a time lag~ \cite{ficekspontaneous2,ficektanaspra2008}.

Here we study the dynamic evolution of classical and quantum correlations corresponding to this peculiar behavior of entanglement. As mentioned before, three main results emerge from our work: (i) The belated appearance of entanglement is preceded by the build up of strong classical and quantum correlations, which so to speak prepare the scenario for the emergence of entanglement; (ii) The interaction between the two initially excited qubits, indirectly through the environment and directly through dipole forces, leads to an entangled state that has extremely small classical correlations; and (iii) although stemming from very different definitions, different quantifiers of quantum correlations coincide within some ranges of time as the system evolves: this is the case for the MID and the quantum discord.

This paper is organized as follows. In Sec. \ref{sec:Quantum and Classical Correlations}. we review the main quantifiers of quantum and classical correlations, with special attention to quantum discord and the ``Measure-Induced Disturbance" (MID) introduced by Luo \cite{luomid}. In Sec. \ref{sec:Dynamics of correlations and entanglement}. we present the master equation that describes the interaction of two qubits with a common environment and the main results concerning the dynamical behavior of several quantities that have been introduced to characterize entanglement and quantum correlations: concurrence, quantum mutual information, quantum discord, and measure-induced disturbance. We also consider in the same section the evolution of the classical correlations. The conclusions are presented in Sec. \ref{sec:conclusions}, while the Appendices contain detailed demonstrations of some of the assertions made in the article.

\section{Quantum and Classical Correlations}
\label{sec:Quantum and Classical Correlations}
\subsection{Quantum discord}
The total classical correlations between two random variables X,Y are given by the Classical Mutual Information:
\begin{equation}
I(X:Y)=H(X)+H(Y)-H(X,Y),
\label{infoclassica1}
\end{equation}
where $
\displaystyle
H(X)=-\sum_{x} p_{(X=x)} \log p_{(X=x)},$
is the Shannon entropy, with the probability distributions calculated from the joint one, $p_{X,Y}$:
$\displaystyle
p_{X}=\sum_{y} p_{X,Y=y}$,
$\displaystyle
p_{Y}=\sum_{x} p_{X=x,Y},$
and $\displaystyle H(X:Y)=-\sum_{x,y}p_{x,y} \log p_{x,y}$ is the joint entropy with $p_{x,y}$ the probability of both outcomes $x$ and $y$ happening.

Using the Bayes rule, $p_{X|Y=y}=p_{X,Y=y}/p_{Y=y}$, where $p_{X|Y=y}$ is the conditional probability that the event $X$ occurs once the event $Y=y$ has already occurred, the Classical Mutual Information can be expressed equivalently as:
\begin{equation}
J(X:Y)=H(X)-H(X|Y),
\label{infoclassica2}
\end{equation}
where  $H(X|Y)=\sum_{y} p_{Y=y} H(X|Y=y)$ is the conditional entropy of $X$ given $Y$, with $\displaystyle H(X|Y=y)=-\sum_{x}p_{x|y}\log p_{x|y}.$
In this form, it becomes clear that the Classical Mutual Information describes the difference in the ignorance about the subsystem $X$ before and after performing a measurement on subsystem $Y$.

Ollivier and Zurek generalized these two equivalent expressions of the Classical Mutual Information to quantum systems. Equation (\ref{infoclassica1}) is easily generalized by replacing the probability distributions by density matrices and the Shannon entropy by the Von Neumann entropy $S(\rho)= - {\rm Tr} (\rho \log_{2}\rho)$, thus obtaining:
\begin{equation}
I(\rho_{A,B})=S(\rho_{A})+S(\rho_{B})-S(\rho_{A,B}).
\label{infoquantica1}
\end{equation}
It is not straightforward to generalize Eq.~(\ref{infoclassica2}), since the definition of the conditional entropy $S(\rho_{A}|\rho_{B})$ involves specifying the state of the subsystem $A$, knowing the state of the subsystem $B$. In quantum theory, this implies that a measurement must be applied to subsystem $B$. The generalization proposed by Ollivier and Zurek is done by assuming a complete unidimensional projective measurement made on system $B$, corresponding to the projectors $\{\Pi_{j}^{B}\}$, such  that $\sum_{j}\Pi_{j}^{B}=1$. The state of $A$ after this measurement is implemented is $\rho_{A|\Pi_{j}^{B}}=\Pi_{j}^{B}\rho_{A,B}\Pi_{j}^{B}/Tr(\Pi_{j}^{B}\rho_{A,B})$, with probability $p_{j}=Tr(\Pi_{j}^{B}\rho_{A,B})$. Then, the conditional entropy is defined as $S(\rho_{A}|\{\Pi_{j}^{B}\})=\displaystyle \sum_{j}p_{j}S(\rho_{A|\Pi_{j}^{B}})$, and the quantum equation corresponding to Eq.~(\ref{infoclassica2}) can be expressed as 
\begin{equation}
J(\rho_{A,B})_{\{\Pi_{j}^{B}\}}=S(\rho_{A})-S(\rho_{A}|\{\Pi_{j}^{B}\}).
\label{infoquantica2}
\end{equation}
The quantum discord is the minimum of the difference between Eq.~(\ref{infoquantica1}) and Eq.~(\ref{infoquantica2}): $D(\rho_{A,B})= \displaystyle \min_{\{\Pi_{j}^{B}\}} (I(\rho_{A,B})-J(\rho_{A,B})_{\{\Pi_{j}^{B}\}}).$ 
The quantity $J(\rho_{A,B})_{\{\Pi_{j}^{B}\}}$ is the information gained about system A when measurements are performed on system B. Through the process of minimization over all possible measurements on system B we search the measurement that disturbs least the total quantum system and allows to obtain as much information as possible from system A.
Ollivier and Zurek demonstrated that
\begin{eqnarray}\label{nonpert}
D(\rho_{A,B})_{\{\Pi_{j}^{B}\}}=0&\Leftrightarrow& \rho_{A,B}=
\sum_{j}\Pi_{j}^{B} \rho_{A,B}\Pi_{j}^{B}, 
\end{eqnarray}
that is, a zero quantum discord implies that the complete quantum system is not disturbed by the measurement. Furthermore, the information about the system $A$ is not perturbed by the measurement of system $B$. However, a nonzero quantum discord implies that the measurement disturbs the state and part of the information about system $A$ that exists in the correlations between the subsystems $A$ and $B$ is lost.
Equation (\ref{nonpert}) shows that the concept of quantum discord admits a simple physical explanation. However, this measure has an annoying feature: it is asymmetric under  exchange of systems $A$ and $B$, which should not be expected from a quantifier of quantum correlations. 
This is made very clear by going back to the example given by Eq.~(\ref{quantumcor}). 
For any complete one-dimensional projective measurement performed on the second qubit, the quantum state of the first one is disturbed and therefore information about it is lost. The quantum discord for this case is different from zero. But if we perform the measurements on the first qubit, the quantum state is not perturbed and all the information on the second qubit, initially present in the state, is recovered. In this case the quantum discord would be zero.
This example also highlights the difference between separability and classicality, and clearly associates the existence of quantum correlations in a separable state with the presence of nonorthogonal states in one of the subsystems, as already mentioned by \cite{poweronequbit}. 
As shown in \cite{olliverzurek}, the quantum discord is always greater or equal than zero, and is zero if and only if the state has only classical correlations.

\subsection{Measurement-Induced Disturbance}
Given any bipartite state $\rho$, one may associate with it, by means of local measurements, another state,  interpreted as the classical part of the former. The quantum correlations present in the state $\rho$ are then determined by quantifying the difference between these two states.
Consider any complete set of one-dimensional orthogonal projections ${\Pi_{i}^{a}}$, ${\Pi_{j}^{b}}$ acting on each party \textit{a} and \textit{b}. The state after the measurement is: $\displaystyle \Pi(\rho)=\sum_{i,j}(\Pi_{i}^{a}\otimes\Pi_{j}^{b})\rho (\Pi_{i}^{a}\otimes\Pi_{j}^{b}).$ If for some measurement $\Pi(\rho)=\rho$, then the state is called classsical state, otherwise the state is truly quantum. Luo demonstrated in \cite{luomid}  that, if $\rho$ is classical, $\{\Pi_{i}^{a}\}$ , $\{\Pi_{j}^{b}\}$ and $\{\Pi_{i}^{a}\otimes\Pi_{j}^{b}\}$ are the eigenprojectors of $\rho_{a}={\rm Tr}_{b}\rho$, $\rho_{b}={\rm Tr}_{a}\rho$ and $\rho$ respectively. This implies that the definition of classical states is unambiguous, since there is a unique measurement that leads to $\Pi(\rho)=\rho$.

When a complete set of projective measurements $\{\Pi_{i}\}$ is performed on a system with state described by the density matrix $\rho$, the entropy of the final state $\Pi(\rho)=\sum_{i}\Pi_{i}\rho\Pi_{i}$ is greater or equal than the entropy of the initial state, $S(\rho)\leq S(\Pi(\rho))$, and equality is attained only when the projective measurements are the eigenprojectors of the matrix $\rho$ \cite{nielsenchuang}. 
In order to quantify the quantum correlations in a quantum state $\rho$, Luo \cite{luomid} chose the measurement $\Pi$ induced by the eigenprojectors of the reduced subsystems. With that choice, the reduced states remain invariant, and then also the corresponding entropies: the information on each subsystem is not changed and, in that sense, this measurement is the least disturbing. 

The corresponding measure of quantum correlations, named ``measurement-induced disturbance",  is defined as: ${\rm MID}(\rho)=I(\rho)-I(\Pi(\rho))$, where $I$ is the quantum mutual information and $\Pi={\Pi_{i}^{a}\otimes\Pi_{j}^{b}}$, with $\Pi_{i}^{a}$ and $\Pi_{j}^{b}$ the eigenprojectors of the reduced subsystems, $\rho_{a}={\rm Tr}_{b}(\rho)$ and $\rho_{b}={\rm Tr}_{a}(\rho)$, respectively. In particular when $\rho$ is a pure bipartite state, ${\rm MID}(\rho)=D(\rho)=S(\rho_{a})$: this measure coincides with the quantum discord, and is equal to the entropy of the reduced system, which is a measure of entanglement.
The MID presents a great advantage with respect to the quantum discord: it is easily calculable, since it does not involve any minimization, one has only to find the eigenvectors of the density matrices corresponding to the subsystems. However, it is not applicable in all cases as Wu \textit{et al.} \cite{Molmer} have remarked: when the local density matrices have degenerate eigenvalues, the MID is not uniquely defined.

\subsection{Classical Correlations}

A quantifier for classical correlations was proposed by Henderson and Vedral \cite{HendVedral}. The Classical Mutual Information given by Eq.~(\ref{infoclassica2}) was generalized to quantum mechanics by replacing the Shannon entropy by the Von Neumann entropy and the classical probability distributions by density matrices. The corresponding quantifier is given by:
\begin{equation}
 C_{A}(\rho_{AB})= \max_{{A_{i}}}(S(\rho_{B})-\sum_{i}p_{i}S(\rho_{B}^{i})),
\label{classicalcorr}
\end{equation}
where
$\rho_{B}^{i}=Tr_{a}(A_{i}\rho_{AB}A_{i}^{\dagger})/Tr(A_{i}\rho_{AB}A_{i}^{\dagger})$
is the state of the subsystem $B$ after performing the POVM $A_{i}$ on $A$. 

Hamieh \textit{et al.}  showed in \cite{hamieh} that, for the case of two qubits, the POVM that maximizes Eq.~(\ref{classicalcorr}) is a complete set of unidimensional orthogonal projectors.
With this result it is easy to see that $I(\rho_{AB})=D(\rho_{AB})+C(\rho_{AB})$, thus the total correlations of the system, quantified by the Quantum Mutual Information (\ref{infoquantica1}), are separated in quantum correlations (measured by the quantum discord) and classical correlations.
One should note that this definition also presents the problem of asymmetry: the quantification of classical correlations depends on which subsystem is measured.

Wu \textit{et al.}  \cite{Molmer} proposed an alternative definition for a quantifier of quantum correlations, which is symmetric and, as opposed to MID, unique even when the states of the subsystems have degenerate eigenvalues. They proposed to quantify the classical correlations between two systems $A$ and $B$ by performing a POVM locally on each subsystem. With the detection records they calculate then the Classical Mutual Information for the chosen POVM.

They define $I_{c_{max}}(\rho_{AB})$ as the maximal classical mutual information available over all choices of possibles POVM's. In order to quantify the quantum correlations in a bipartite state, they define $Q(\rho_{AB})=I(\rho_{AB})-I_{c_{max}}(\rho_{AB})$. When $\rho_{AB}$ is a pure state $Q(\rho_{AB})=S(\rho_{A})=D(\rho_{AB})={\rm MID}(\rho_{AB})$, and in the case of a general mixed state, ${\rm MID}(\rho_{AB})\ge Q(\rho_{AB})\ge D(\rho_{AB}).$

In this work we refrain from studying the dynamical behavior of the quantity $Q$, because it requires an optimization over all possible measurements, and an analytical expression for it is still unknown. 

\subsection{Entanglement}

For the quantification of entanglement we use the  \textit{Concurrence} \cite{wootters}, defined as:
 \begin{equation}
{\cal C} = {\rm max}\{0,\Lambda\},
\label{concurrence}
\end{equation}
where:
\begin{equation}
\Lambda=\sqrt{\lambda_{1}}-\sqrt{\lambda_{2}}-\sqrt{\lambda_{3}}-\sqrt{\lambda_{4}}\,,
\label{autovalores}
\end{equation}
$\lambda_{i}$ being the eigenvalues in decreasing order of the matrix:
\begin{equation}
\rho(\sigma_{y}\otimes\sigma_{y})\rho^{*}(\sigma_{y}\otimes\sigma_{y}),
\label{lamatriz}
\end{equation}
where $\rho$ is the two-qubit density matrix, $\sigma_{y}$
is the second Pauli matrix and the conjugation is performed in the
computational basis. Concurrence
ranges from 0, which corresponds to a separable state, to 1, which corresponds to a maximally entangled state.
\section{Dynamics of correlations and entanglement}
\label{sec:Dynamics of correlations and entanglement}

\subsection{Theoretical Model}

We consider a system of two identical qubits interacting with all modes of the electromagnetic field, assumed in the vacuum state. The state of each qubit is represented in the basis $\{{\ket{e},\ket{g}}\}$ ($e=$ excited, $g=$ ground state). The two qubits interact via dipole forces, associated to the dipole transition moments $\mu$.The total Hamiltonian of the atoms plus the electromagnetic field, in the electric dipole approximation, is
\begin{align*}
\hat{H}&= \sum_{i}^{2}\hbar\omega_{i}S_{i}^{z}+\sum_{\vec{k}}\hbar\omega_{k}\hat{a}_{\vec{k}s}^{\dagger}\hat{a}_{\vec{k}s}\\
&-i\hbar\sum_{\vec{k}s}\sum_{i=1}^{2}\bigg[ \vec{\mu}.\vec{g}_{\vec{k}s}(\vec{r}_{i})(S_{i}^{+}+S_{i}^{-})\hat{a}_{\vec{k}s}-H.c\bigg],
\end{align*}
where $S_{i}^{+}=\ket{e_{i}}\bra{g_{i}}$ and $S_{i}^{-}=\ket{g_{i}}\bra{e_{i}}$ are the ladder operators, $S_{i}^{z}=\ket{e_{i}}\bra{e_{i}}-\ket{g_{i}}\bra{g_{i}}$ is the energy operator of the ith qubit, $\omega_{i}$ are the transition frequencies (in what follows we will consider all frequencies equal, $\omega_{i}=\omega_{0}$), $\hat{a}_{\vec{k}s}$ and $\hat{a}_{\vec{k}s}^{\dagger}$ the annihilation and creation operators corresponding to the field mode $\vec{k}s$, with wave vector $\vec{k}$, frequency $\omega_{k}$ and index of polarization $s$.
$\displaystyle \vec{g}_{\vec{k}s}(\vec{r}_{i})=\bigg( \frac{\omega_{k}}{2\varepsilon_{0}\hbar V}\bigg)^{1/2}\bar{e}_{\vec{k}s}e^{i\vec{k}.\vec{r}_{i}}$ is the coupling constant, $\vec{r}_{i}$ is the position of the ith qubit, $V$ is the normalization volume, and $\bar{e}_{\vec{k}s}$ is the unit polarization vector of the field. When the length of the system is small compared to the radiated wavelength $(\omega_{0}/c)$, we may neglect the spatial variation of $\vec{g}_{\vec{k}s}(\vec{r}_{i})$, so that, in the rotating-wave approximation, the Hamiltonian reduces to \cite{agarwal}:
\begin{align*}
\hat{H}=\hbar\omega_{0}S^{z}+\sum_{\vec{k}s}\omega_{k}a_{\vec{k}s}^{\dagger}a_{\vec{k}s}-i\hbar\sum_{\vec{k}s}[\vec{\mu}.\vec{g}_{\vec{k}s}S^{+}\hat{a}_{\vec{k}s}-H.c],
\end{align*}
where $S^{\pm}$ and $S^{z}$ are collective spin operators defined by: $\displaystyle S^{\pm}=\sum_{i}S_{i}^{\pm}$ and $\displaystyle S^{z}=\sum_{i}S_{i}^{z}$.

This Hamiltonian describes the Dicke model. The other limit, when the length of the system is much greater than the resonant wavelength, is easily obtained from the master equation approach that we considerer in the following.

The dynamical evolution of the qubit system is given by the following master equation \cite{ficekspontaneous2}:
\begin{eqnarray}
\frac{\partial {\rho}(\tau)}{\partial t}&=&
-i \omega_0\sum_{i=1}^{2}[S_{i}^{z},\rho]-
i \Omega_{12}\sum_{i\ne j}^{2}[S_{i}^{+}S_{j}^{-},\rho] \nonumber \\
&&-\frac{1}{2} \sum_{i,j=1}^{2} \Gamma_{ij}(\rho S_{i}^{+}S_{j}^{-}+S_{i}^{+}S_{j}^{-}\rho-2S_{j}^{-}\rho S_{i}^{+})\,,
\nonumber\\
\label{mastereq}
\end{eqnarray}
where $\Gamma_{ii}\equiv\Gamma$  are the spontaneous emission rates of the qubits, assumed to be identical, and $\Gamma_{12}=\Gamma_{21}$ and
$\Omega_{12}$ are respectively the collective damping and the dipole-dipole interaction defined by
\begin{align}\label{gamma}
\Gamma_{12}&= \frac{3}{2} \Gamma 
\Bigg\{ 
\Big[ 1 - \left( \hat{\mu} . \hat{r}_{12} \right) ^{2} \Big] 
\frac{\sin(k_{0}r_{12})}{k_{0}r_{12}} \nonumber \\
&+\Big[ 1 - 3 \left( \hat{\mu} . \hat{r}_{12} \right)^{2} \Big]\bigg[ \frac{\cos(k_{0}r_{ij})}{(k_{0}r_{12})^2}- \frac{\sin(k_{0}r_{12})}{(k_{0}r_{12})^3} \bigg] \Bigg\}, 
\end{align}
and
\begin{align}\label{Om}
\Omega_{12}&= \frac{3}{4}\Gamma \Bigg\{
-\Big[1- (\hat{\mu}.\hat{r}_{12})^{2}\Big] \frac{\cos(k_{0}r_{12})}{k_{0}r_{12}}\nonumber \\
&+\Big[1-3(\hat{\mu}.\hat{r}_{12})^{2}\Big]
\bigg[\frac{\sin(k_{0}r_{12})}{(k_{0}r_{12})^{2}}+\frac{\cos(k_{0}r_{12})}{(k_{0}r_{12})^{3}}                \bigg] \Bigg\}, 
\end{align}
where $k_{0}=\omega_{0}/c$, $r_{12}=|\mathbf{r}_{1}-\mathbf{r}_{2}|$ is the distance between the qubits, and $\mathbf{\mu}$ as before is the dipole transition moment. It follows from Eq.~(\ref{gamma}) that $\gamma\equiv \Gamma_{12}/\Gamma\le1$.

When the distance between the qubits is much greater than the resonant wavelenght, it is easy to see that  $\Omega_{ij}\rightarrow 0$ and $\Gamma_{ij}\rightarrow 0, i\neq j$, so that the master equation becomes
\begin{align*}
\frac{\partial {\rho}(t)}{\partial t}&=
-i\omega_0 \sum_{i=1}^{2}[S_{i}^{z},\rho]\\
&-\frac{1}{2}\Gamma \sum_{i}^{2} (\rho S_{i}^{+}S_{i}^{-}+S_{i}^{+}S_{i}^{-}\rho-2S_{i}^{-}\rho S_{i}^{+})\,,
\label{mastereqSep}
\end{align*}
which corresponds to the case of independent environments.

\subsection{Results}

\begin{figure}[b]
\begin{center}       
\includegraphics[scale=1.4]{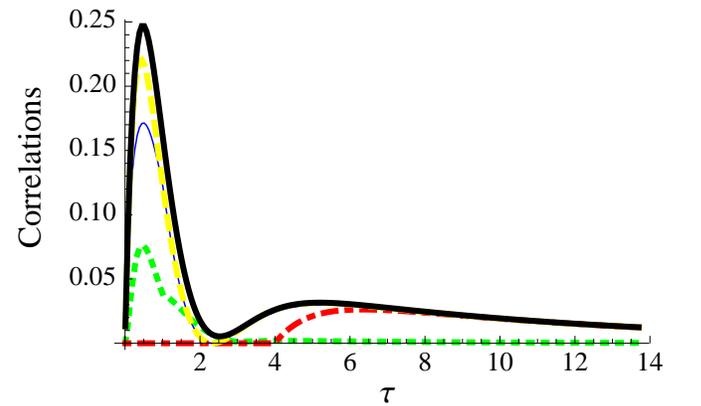}
\end{center}
\captionsetup{justification=centerlast}
\caption{\footnotesize (Color online) Evolution of quantum correlations: The thick black line is the quantum mutual information, the thin blue line is the quantum discord, the dashed yellow line is the measured-induced disturbance, the green dotted line is the classical correlation, and the red dotted-dashed line is the concurrence. ($r_{12}=\lambda/8, \gamma=0.8806$) } \label{correl} 
\end{figure}

In order to study the dynamical evolution of quantum and classical correlations, the master equation (\ref{mastereq}) was solved analytically; we refrain from showing the details of the solution here since our results coincide with those in Ref.~\cite{ficekspontaneous2}. From the analytical solution,  we calculated the relevant quantities for our purposes: concurrence, quantum discord, MID and classical correlation, for several initial states.\\
The dynamical evolution of these quantities when both qubits are initially excited, interacting with a common reservoir and with each other via dipole-dipole interaction, is shown in Fig.~\ref{correl}. 
We see that initially all correlations, with the exception of the concurrence that remains zero, rise to  a maximum value and then decrease. One should also note that, for a finite period of time, the MID and the quantum discord coincide. This can be understood by using the results in \cite{caldeira09}; for a subset of density matrices with structure X,
\begin{eqnarray}
\rho=\left(\begin{array}{cccc}
           a&0&0&\omega\\
           0&b&z&0\\
            0&z&b&0\\
            \omega&0&0&d
           \end{array}\right)\,, \nonumber
           \label{rhocaldeira}
\end{eqnarray}
the quantum discord $D(\rho)$ is given by: $D(\rho)= {\rm min} \{D_{1},D_{2}\}$, where 
\begin{eqnarray}
D_{1}&=& S(\rho^{A})-S(\rho^{AB})-a\log_{2}\bigg(\frac{a}{a+b}\bigg)-
b\log_{2}\bigg(\frac{b}{a+b}\bigg) \nonumber \\
&-&d\log_{2}\bigg(\frac{d}{b+d}\bigg)-b\log_{2}\bigg(\frac{b}{b+d}\bigg),
\label{d1}
\end{eqnarray}
and  
\begin{eqnarray}
D_{2}&=
S(\rho^{A})-S(\rho^{AB})-\frac{1}{2} (1+\alpha)\log_{2}\bigg(\frac{1}{2} (1+\alpha)   \bigg) \nonumber \\
&-\frac{1}{2}(1-\alpha)\log_{2}\bigg(\frac{1}{2}(1-\alpha)\bigg),
\label{d2}
\end{eqnarray}
with $\alpha^{2}=(a-d)^{2}+4|z+\omega|^{2}.$ \\

The density matrix describing the dynamical evolution of the two-qubit system here considered, when both qubits are initially excited, has the same form as the one in  Eq.~(\ref{rhocaldeira}). Indeed, we get
\begin{eqnarray}
\rho(\tau)=\left(\begin{array}{cccc}
           a(\tau)&0&0&0\\
           0&b(\tau)&c(\tau)&0\\
            0&c(\tau)&b(\tau)&0\\
            0&0&0&1-a(\tau)-2b(\tau)
           \end{array}\right)\,,
           \label{rhonossa}
\end{eqnarray}
where
\begin{eqnarray}
a(\tau)&=&e^{-2\tau} \nonumber \\
b(\tau)&=&\frac{[-e^{-2\tau}+e^{-(1-\gamma)\tau}](1 -\gamma)}{2(1 + \gamma)} \nonumber \\
&+&\frac{[-e^{-2\tau}+e^{-(1+\gamma)\tau}](1 + \gamma)}{2(1 -\gamma)} \nonumber \\
c(\tau)&=&-\frac{[-e^{-2\tau}+e^{-(1-\gamma)\tau}](1 -\gamma)}{2(1 + \gamma)}\nonumber \\
&+&\frac{[-e^{-2\tau}+e^{-(1+\gamma)\tau}](1 + \gamma)}{2(1 -\gamma)}\,,
\label{coeff}
\end{eqnarray}
with
\begin{equation}
\tau=\Gamma t\,,\quad \gamma=\Gamma_{12}/\Gamma\,.
\end{equation}

As a function of the renormalized time $\tau$, these expressions depend on a single parameter $\gamma$. It is easy to show from thie above equations that  the analytic expression for the MID coincides with the expression for $D_{1}$ (see Appendix A), so whenever $D_{1}<D_{2}$, the MID coincides with the quantum discord. This condition is fulfilled for a finite time domain, that depends on the parameter $\gamma$ as shown in Fig.~\ref{d1d2}.  

At a later time $t_{e}$, the system becomes entangled and the concurrence starts increasing. This feature was previously found by Ficek and Tana\'s in \cite{ficekspontaneous2,ficektanaspra2008}. They showed that, for  two  qubits initially in their excited states, under the conditions considered in this paper (interacting with a common environment and through dipole forces),  and for small distance $r_{12}$ between them (as compared to the resonance wavelength), entanglement appears only after some finite period of time. This feature is independent of the values of the parameters, as shown in Appendix B. 
After the time interval where MID and quantum discord match, the numerical value of the quantum discord remains close to the MID, but always a little lower. On the other hand, after the emergence of entanglement, the difference between MID and Concurrence continuously decreases, going to zero when both vanish at infinite times.

Classical correlations are smaller than quantum correlations for most of the dynamic evolution of the state, a result that was previously
found in \cite{luo2}. After they reach their maximum value, they go to zero asymptotically. However, for most of the time of the evolution, they have a very small value, about a hundred times smaller than the concurrence. 
This behavior is quite general: we have demonstrated it for several values of the  parameter $\Gamma_{12}$, as shown in Appendix C.
In Ref.~\cite{vedral08}, it was shown that only with three or more qubits can one have states with genuine quantum correlations but no classical correlations. Our example does not contradict that general result, since for the two-qubit system here considered the classical correlations, as defined in \cite{Henderson02}, are indeed different from zero, but it also shows that the same correlations, even though not vanishing, may be much smaller than the quantum correlations, even for a two-qubit system. 

Ficek and Tana\'s also showed in Ref.~\cite{ficektanaspra2008} that, for the Dicke model (evolution of two two-level atoms in a common reservoir with no dipole-dipole interaction), two qubits initially excited do not get entangled. A question naturally arises as to whether that is the case for the other quantum correlations in the Dicke model. 
This question is answered by taking the limit $r_{12}\rightarrow 0$ in the expressions for $\Gamma_{12}$ and $\Omega_{12}$. In that case,  $\Gamma_{12}\rightarrow\Gamma$, so that $\gamma\rightarrow 1$, and $\Omega_{12}\rightarrow[3\Gamma/4(k_{0}r_{12})^3][1-3(\vec{\mu}.\vec{r_{12}})^{2}]$. Although
$\Omega_{12}$ diverges in this limit, Eq.~(\ref{coeff}) shows that the evolution of the initial quantum state considered here is independent of the dipole-dipole interaction. Therefore, in this limit we do recover, for the considered initial state, the evolution predicted by the Dicke model. 
The density matrix for this model is easily obtained from Eqs.~(\ref{rhonossa}) and ~(\ref{coeff}) taking the limit $\gamma\rightarrow 1$:
\begin{eqnarray}
\rho(\tau)=\left(\begin{array}{cccc}
           a(\tau)&0&0&0\\
           0&b(\tau)&b(\tau)&0\\
            0&b(\tau)&b(\tau)&0\\
            0&0&0&1-a(\tau)-2b(\tau)
           \end{array}\right)\,,
           \label{rhodicke}
\end{eqnarray}
where
\begin{eqnarray}
a(\tau)&=&e^{-2\tau} \nonumber \\
b(\tau)&=& e^{-2\tau}\, \tau\,.
\label{coeffdicke}
\end{eqnarray}

\begin{figure}[t]
\begin{center}
\includegraphics[scale=1.6]{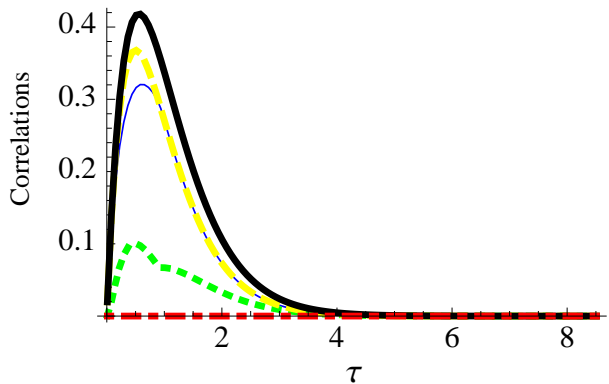}
\end{center}
\captionsetup{justification=centerlast}
\caption{\footnotesize
(Color online) Evolution of the quantum correlations for the Dicke model: the thick black line is the Quantum Mutual Information, the thin blue line is the quantum discord, the dashed yellow line is the MID, the green dotted line is the Classical Correlation and the red dotted-dashed line is the Concurrence.}
\label{correldicke}
\end{figure}

Figure~\ref{correldicke} displays the evolution of all the above-mentioned correlations, within the framework of the Dicke model, when the two qubits are initially excited. 
In this case, the initial behavior of classical and quantum correlations (except entanglement) is the same as in the model that includes dipole-dipole interaction: an initial period of growth to a maximum and then a decay. However, one does not have, as in the former situation, a revival of these correlations. 

\begin{figure}[t]
\begin{centering}
\includegraphics[scale=0.86]{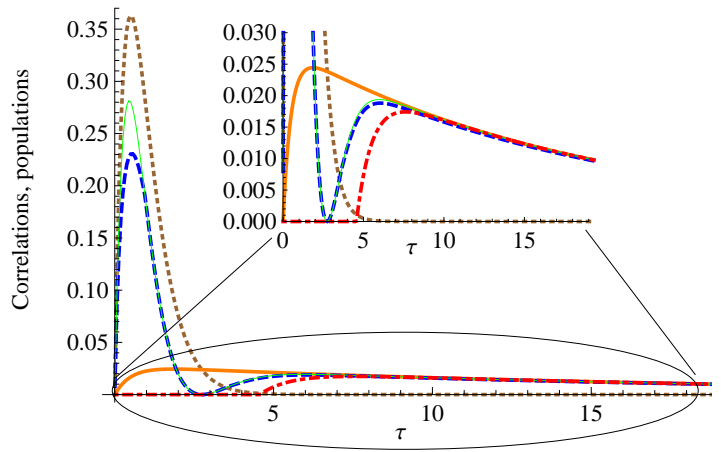}
\end{centering}
\captionsetup{justification=centerlast}
\caption{\footnotesize (Color online) Evolution of the quantum correlations when the dipole-dipole interaction is included: The green solid line is the quantum discord, the blue dashed line is the MID, the red thick dashed-dotted line is the Concurrence, the brown dotted line is the population of the symmetric state and the orange thick line is the population of the antisymmetric state. The inset shows the zone inside the ellipse. ($r_{12}=\lambda/4, \gamma=0.9460$)} \label{decayasimetrico}
\end{figure}

The lack of entanglement for an initially doubly excited two-qubit state $|ee\rangle$ in the absence of dipole-dipole interaction stems from the fact that the singlet state $\frac{1}{\sqrt{2}}(\ket{eg}-\ket{ge})$ does not show up in the evolution, as discussed by Ficek and Tana\'s \cite{ficekspontaneous2}: this is a decoherence-free state in the Dicke model,  and its constant contribution leads to asymptotic entanglement when the initial state is $|eg\rangle$, since this state can be written as a sum of the anti-symmetric singlet $\frac{1}{\sqrt{2}}(\ket{eg}-\ket{ge})$ and the symmetric state $\frac{1}{\sqrt{2}}(\ket{eg}+\ket{ge})$. The dipole-dipole interaction, on the other hand, leads to the emergence of the singlet state for the initial state $|ee\rangle$, to the survival of quantum correlations, and to the belated appearance of entanglement.  
Indeed, the analytical expressions for the MID and quantum discord show that, for $\tau>>1$, the decay time of the symmetric state, both MID and quantum discord decay as $e^{-(1-\gamma)\tau}$, which is the decay law for the population of the singlet state. This is  shown in Figure~\ref{decayasimetrico}, which makes clear the essential role of the singlet state in explaining the remarkable behavior of the entanglement and the quantum correlations in the presence of dipole-dipole interactions.  

When the distance between the qubits  $r_{12}$ becomes much larger than the resonant wavelength, we recover the expected behavior for two initially excited qubits  evolving in independent environments: the state remains separable and no correlations are created between the qubits.

\section{CONCLUSIONS}
\label{sec:conclusions}

In this work, we have studied the subtle dynamics of quantum and classical correlations for two qubits, initially in a pure product state, coupled through dipole forces and interacting with all modes of the electromagnetic reservoir.  
When the two qubits are initially excited, entanglement has a peculiar behavior \cite{ficekspontaneous2,ficektanaspra2008}: it remains zero for a finite-time interval, and then it builds up. Our work is aimed at clarifying what happens during the ``dormant" time, that is, what kind of dynamic changes prepare the system for the late appearance of entanglement.
With this aim, we have analyzed the dynamic behavior of quantifiers of quantum and classical correlations. We show that, for two excited qubits,  the reservoir creates initially classical correlations and quantum correlations between the qubits, which remain in a separable state. These correlations evolve from zero to a maximum value, and then decay. This overall behavior does not depend on the presence of dipole-dipole interactions. However, this interaction is fundamental for keeping up the quantum correlations and building up entanglement after a time lag. It delays the decay of quantum correlations, thus allowing for the build up of entanglement. 

Therefore, in the presence of dipole-dipole interactions, quantum correlations can be considered as precursors of entanglement. In the absence of these interactions, they build up and decay to zero, the state remaining separable for all times. The dipole-dipole interaction helps preserving quantum correlations, and this seems to fire up entanglement in this case.
While entanglement is still zero, there is a time span for which MID and quantum discord coincide. 
Even before entanglement appears, classical correlations become negligible, and remain so throughout the evolution. Therefore, the very evolution of the system generates a peculiar entangled state, with very small classical correlations. 
Even though there is still considerable controversy over the proper definition of quantum correlations and their role in quantum computation, their dynamics under the action of the environment seems to be, in the present context, intimately related to the generation of entanglement. Further studies in this direction might help to elucidate this subtle dynamic connection between quantum correlations and entanglement.

\section{Appendix}
\label{sec:apendice}
\subsection{MID equals quantum discord}
\begin{figure}[h]
\begin{centering}
\includegraphics[scale=0.62]{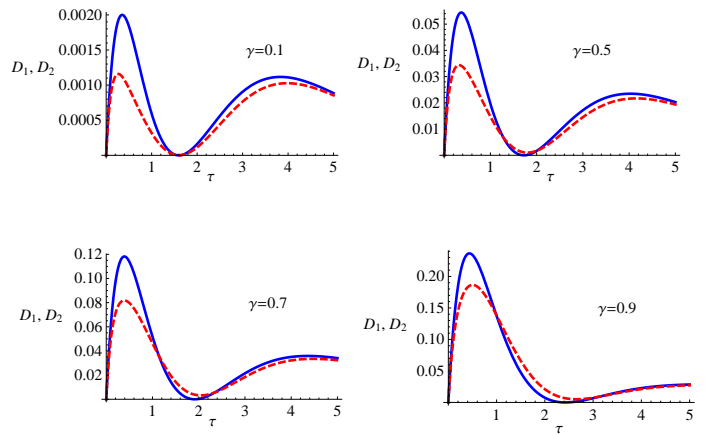}
\end{centering}
\captionsetup{justification=centerlast}
\caption{\footnotesize (Color online) Evolution of the expressions $D_{1}$ (blue line) and $D_{2}$ (red dashed line) for different values of the parameter $\gamma$.} \label{d1d2}
\end{figure}

The density matrix corresponding to the initial state is given by  Eq.~(\ref{rhonossa}). 
To calculate the MID, we apply projective measurements on the density matrix, corresponding to the eigenvectors of the reduced density matrices. After these measurements, which amount to keeping only the diagonal elements in Eq.~(\ref{rhonossa}), the density matrix is:
\begin{eqnarray*}
\rho_{meas}(\tau)=\left(\begin{array}{cccc}
           a(\tau)&0&0&0\\
           0&b(\tau)&0&0\\
            0&0&b(\tau)&0\\
            0&0&0&1-a(\tau)-2b(\tau)
           \end{array}\right)\,.
           \label{medidas}
\end{eqnarray*}
Then 
\begin{align} 
{\rm MID}(\rho)&=-S(\rho)+S(\rho_{meas})=-2b(\tau)\log_{2}[b(\tau)]\nonumber \\
&+[b(\tau)-c(\tau)]\log_{2}[b(\tau)-c(\tau)] \nonumber \\
&+[b(\tau)+c(\tau)]\log_{2}[b(\tau)+c(\tau)],
\label{mididiscordia}
\end{align} 
expression that coincides with $D_{1}$, Eq.~(\ref{d1}), throughout the evolution. According to \cite{caldeira09} the quantum discord is the minimum of $\{D_{1},D_{2}\}$. Therefore, whenever $D_{1} < D_{2}$, the MID coincides with the quantum discord.
The expressions for $D_{1}$ and $D_{2}$ do not allow getting 
an analytical solution for the interval of time where $D_{1} < D_{2}$. Fig.~\ref{d1d2} exhibits the results of numerical calculations made for several values of  $\gamma$, displaying time intervals (reduced to a single point for small $\gamma$) where $D_{1} < D_{2}$.

\begin{figure}[b]
\begin{centering}
\includegraphics[scale=0.89]{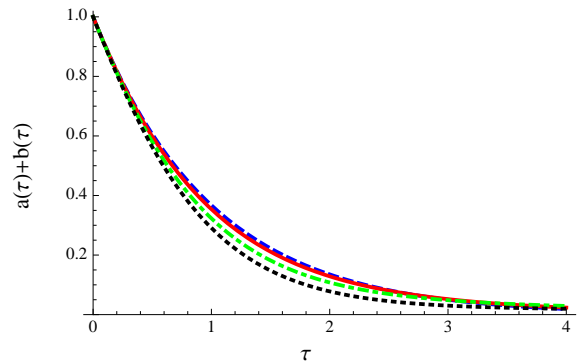}
\end{centering}
\captionsetup{justification=centerlast}
\caption{\footnotesize (Color online) Plots of $a(t)+b(t)$ for $\gamma=$ 0.1 (blue dashed line), 0.4 (red full line), 0.7 (green dashed-dotted line) and 0.9 (black dotted line).} \label{ab}
\end{figure}

As mentioned above, in the calculation of the MID there is an ambiguity when the reduced matrices have degenerate eigenvalues, or equivalently, are multiples of the identity matrix. In this case, any two orthogonal vectors are eigenvectors of the reduced matrices, so there are infinite ways to choose the local projective measurements, which give rise to different values of the MID. Here the reduced density matrices are:
\begin{eqnarray*}
\rho_{a}(\tau)=\rho_{b}(\tau)=\left(\begin{array}{cc}
           a(\tau)+b(\tau)&0\\
           0&1-a(\tau)-b(\tau)
           \end{array}\right)\,,
           \label{reduzidas}
\end{eqnarray*}
and the mathematical condition for the reduced matrices to become proportional to the identity is: 
\begin{equation}
a(\tau)+b(\tau)=\frac{1}{2},
\label{degenerados}
\end{equation}
with
\begin{eqnarray*}
a(\tau)+b(\tau)&=& e^{-2\tau}+ \frac{e^{-\tau}}{(1-\gamma^2)}
\left\{ [\cosh(\gamma\tau)-e^{-\tau}]\right. \\ 
&&\left. \times(1+\gamma^2)-\gamma \sinh (\gamma\tau) \right\}\,.
\end{eqnarray*}
This expression decreases monotonically from its initial unitary value to zero, as $\tau\rightarrow\infty$ -- see Fig.~\ref{ab}.  Therefore, there can only be a single instant of time for which   Eq.~(\ref{degenerados}) holds. 
 At this point, the MID is not uniquely defined and depends on the eigenvectors chosen to make the measurements. However, since, for all other times, the eigenvectors are uniquely defined and coincide with the computational basis, it is natural to choose the same basis for the degeneracy points, in which case the MID does not present discontinuities.

\subsection{Concurrence}

The concurrence $\cal{C}$ is expressed in terms of the square roots of the eigenvalues of the matrix in Eq.~(\ref{lamatriz}): 
\begin{eqnarray}\label{values}
&\{\sqrt{a(\tau)\left[1-a(\tau)-2b(\tau)\right]},\, \sqrt{a(\tau)\left[1-a(\tau)-2b(\tau)\right]},\nonumber\\
&\left|b(\tau)-c(\tau)\right|,\, \left|b(\tau)+c(\tau)\right|\}\,.
\end{eqnarray}
First we will show that $b(\tau)>c(\tau)$. This is immediate since, from Eq.~(\ref{coeff}), 
\begin{equation*}
b(\tau)-c(\tau)=\frac{e^{-\tau}}{1+\gamma}(1-\gamma)(e^{\gamma \tau}-e^{ -\tau})\,.
\end{equation*}
We distinguish three different cases: (1) $\sqrt{a(\tau)\left[1-a(\tau)-2b(\tau)\right]}$ is the largest value in Eq.~(\ref{values}), in which case ${\cal C}=  {\rm max}\{0,-2b(\tau)\}$, (2)  $\left|b(\tau)+c(\tau)\right| $ is the largest value, in which case  $ {\cal C}= {\rm max}\bigg\{0,2\left(c(\tau)-\sqrt{a(\tau)\left[1-a(\tau)-2b(\tau)\right]}\right)\bigg\}$,  and (3) $\left|b(\tau)-c(\tau)\right|$ is the largest value, in which case ${\cal C}= {\rm max}\bigg\{0,-2\left(c(\tau)+\sqrt{a(\tau)\left[1-a(\tau)-2b(\tau)\right]}\right)\bigg\}$.
 
In the first case, it follows from the expression for $b(\tau)$ in Eq.~(\ref{coeff}) and from the inequality  $\cosh(\gamma\tau)-e^{-\tau} > \sinh(\gamma\tau)$ that $b(\tau)>0$ for all $\tau$. Therefore, the concurrence is zero in this case. 

In the second case, we must investigate the behavior of $c(\tau)-\sqrt{a(\tau)[1-a(\tau)-2b(\tau)]}$.  First, we note that this case is realized only in the interval where $c(\tau)>0$, so that  $c(\tau)^2<b(\tau)c(\tau)$. We show now that $b(\tau)c(\tau)\leq a(\tau)(1-a(\tau)-2b(\tau))$, which implies that $c(\tau)<\sqrt{a(\tau)[1-a(\tau)-2b(\tau)]}$. This is equivalent to proving that $\displaystyle f(\tau)\equiv a(\tau)+2b(\tau)+\frac{b(\tau)c(\tau)}{a(\tau)}\leq 1$. Since $f(0)=1$, the equality is verified at $\tau=0$. Furthermore, it is easy to show that $f'(\tau)<0$ for all $\tau$, independently of the value of the parameter. Therefore,  $f(\tau)$ is a monotonically decreasing function of $\tau$ with $f(0)=1$, so $f(\tau)\leq 1$ for all $\tau$, implying that $c(\tau)<\sqrt{a(\tau)[1-a(\tau)-2b(\tau)]}$. 
Thus, we conclude that for this second case the concurrence is also zero.

In the third case, we must investigate the behavior of $g(\tau)\equiv c(\tau)+\sqrt{a(\tau)[1-a(\tau)-2b(\tau)]}$. 
Because the equation $g(\tau)=0$ contains the time in a non-algebraic expression, it is not possible to determine analytically the time when the state becomes entangled; however we can show, through some simple mathematical arguments, that, whatever the value of the parameter, this time always exists. This requires to show that there is a time interval for which two conditions are satisfied: (i) $g(\tau)<0$;  and (ii) $\left|b(\tau)-c(\tau)\right|$ is the largest value in Eq.~(\ref{values})

Condition (i) is demonstrated by noting that $\displaystyle \lim_{\tau\rightarrow 0} g(\tau)=\lim_{\tau\rightarrow 0}\sqrt{\tau \frac{1+\gamma^2}{1-\gamma^2} }=0^{+}$, and  $\displaystyle \lim_{\tau\rightarrow +\infty} g(\tau)=\lim_{\tau\rightarrow +\infty}-\frac{(1 -\gamma)}{2(1+\gamma)}e^{-(1 -\gamma)\tau}=0^-$. Then between zero and infinite there exists an instant of time $\tau_{e}$ for which $g(\tau)=0$. This implies the existence of a time interval from $\tau_e$ to infinity where $g(\tau)$ is negative, which would imply entanglement, as long as  condition (ii) is satisfied within at least part of this time interval.

We show now that condition (ii) holds for sufficiently large $\tau$. This follows from $\displaystyle \lim_{\tau\rightarrow +\infty}c(\tau)=\lim_{\tau\rightarrow +\infty}-\frac{(1-\gamma)}{2(1+\gamma)}e^{-(1-\gamma)\tau}=0^-$, which implies that $c(\tau)<0$ for sufficiently large times. Therefore, for times sufficiently large, one has $g(\tau)<0$ and $\left|b(\tau)-c(\tau)\right|>\left|b(\tau)+c(\tau)\right|$. It remains to show that, for sufficiently large times,  $\left|b(\tau)-c(\tau)\right|>\sqrt{a(\tau)\left[1-a(\tau)-2b(\tau)\right]} $. This is shown by noting that there is a time $\tau_{e}'$ such that, for all $\tau>\tau_{e}'$, $\displaystyle e^{\gamma \tau}-e^{-\tau} > \frac{(1+\gamma)}{(1-\gamma)}$, which is immediate. Therefore, for sufficiently large times,  both conditions (i) and (ii) above are fulfilled, and the state becomes entangled.

\subsection{Classical Correlations}
\begin{figure}[htp]
\subfloat[]{
\label{cc01}
\includegraphics[scale=0.5]{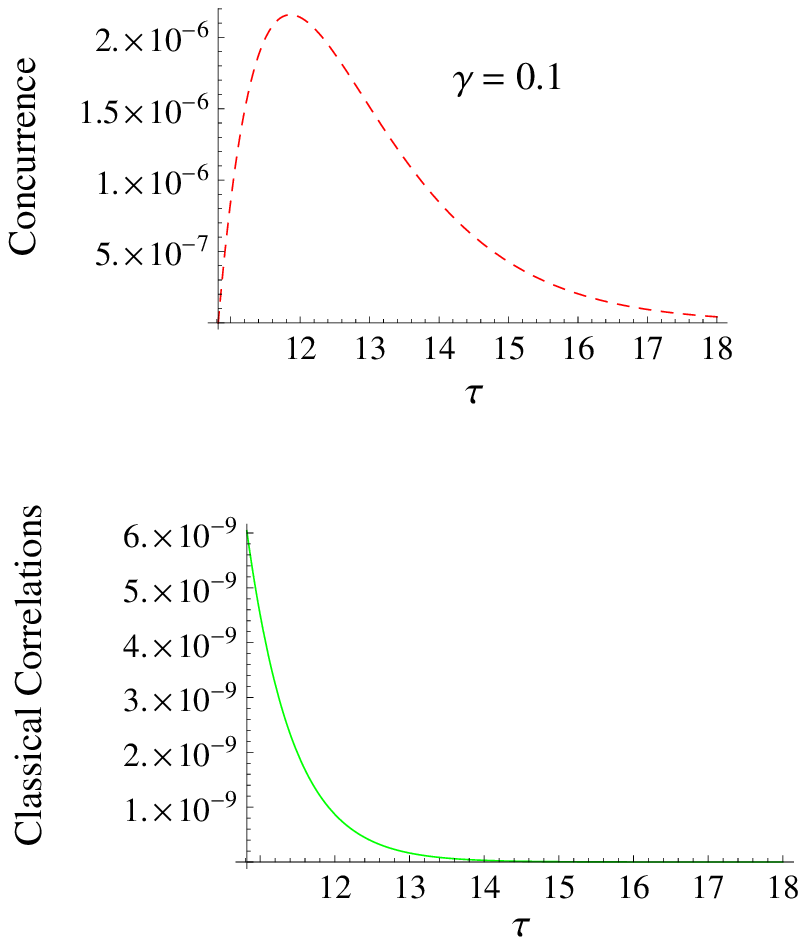}}
\hspace{10pt} 
\subfloat[]{
\label{cc05}
\includegraphics[scale=0.5]{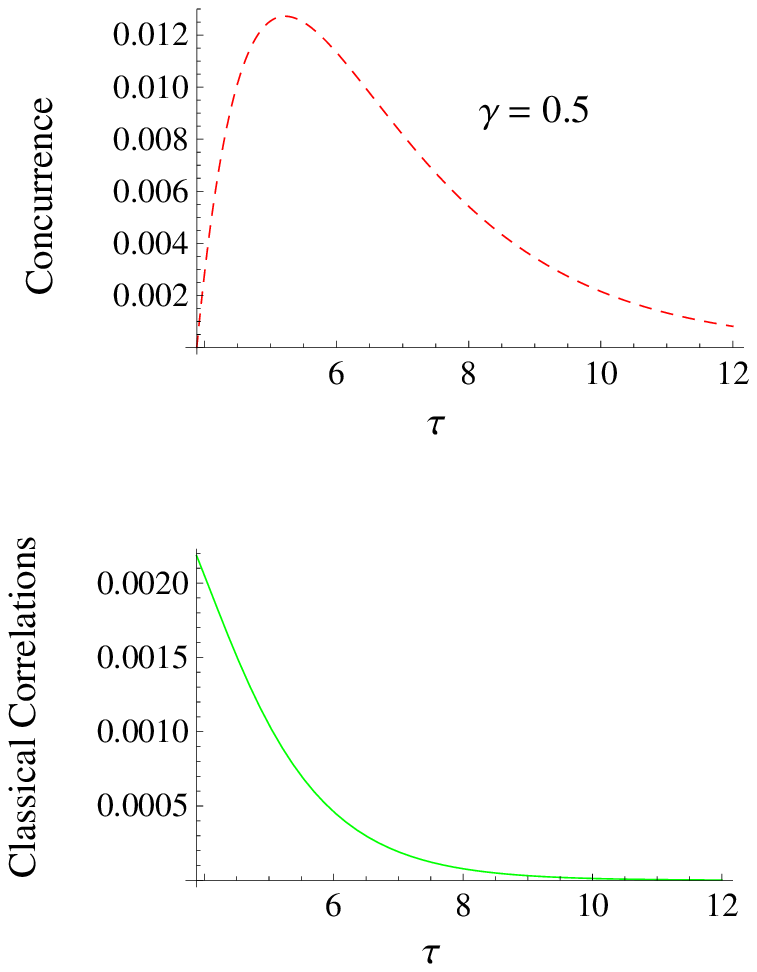}
} \\[10pt]
\subfloat[]{
\label{cc07}
\includegraphics[scale=0.55]{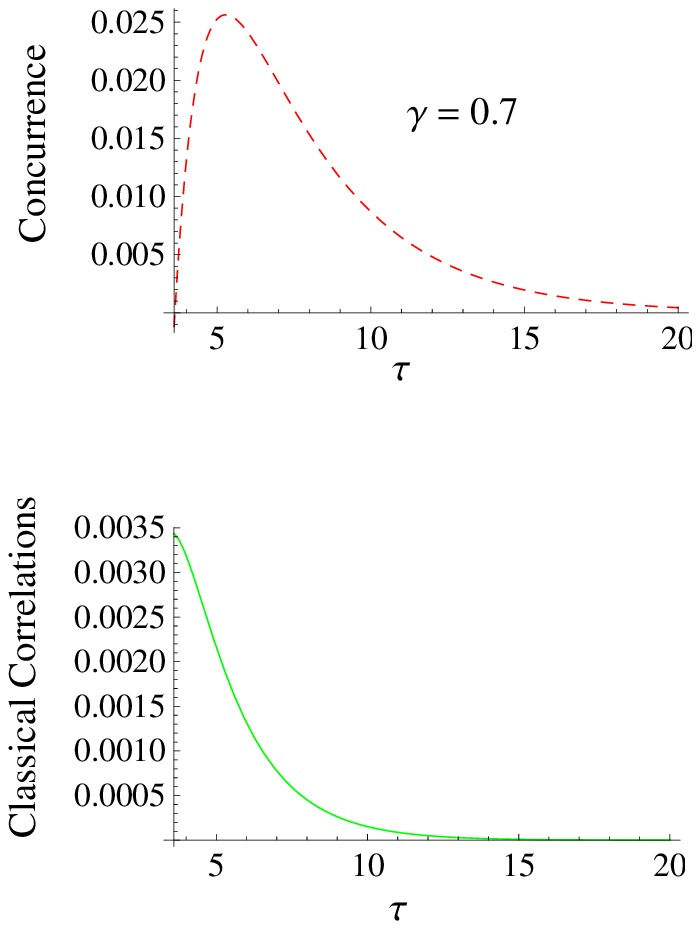}
}
\hspace{10pt}
\subfloat[]{
\includegraphics[scale=0.55]{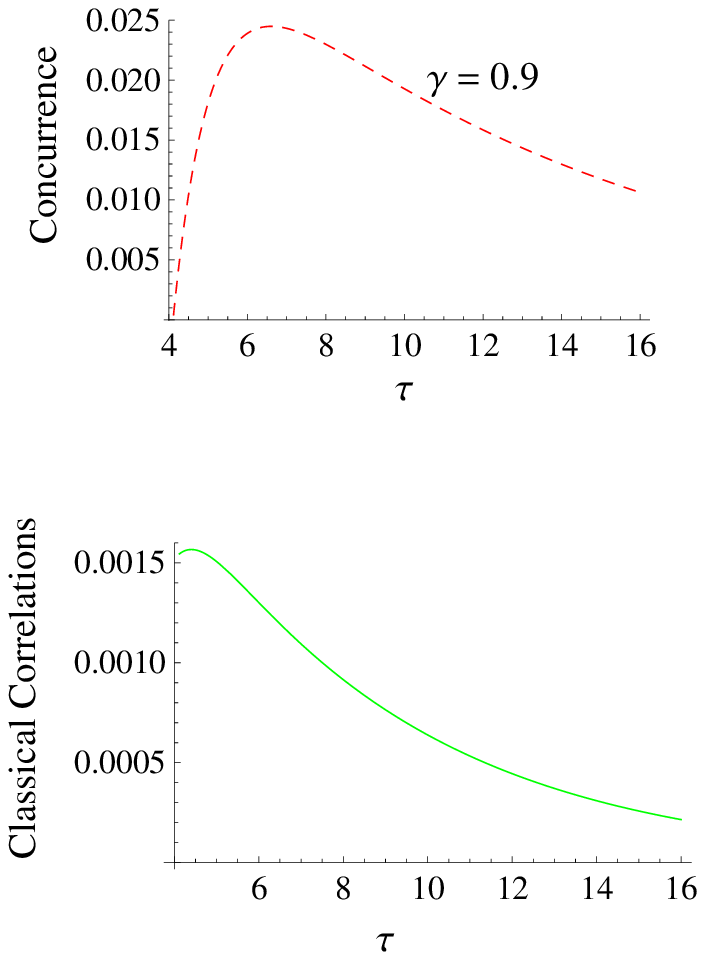}
}
\captionsetup{justification=centerlast}
\caption{\footnotesize (Color online) Evolution of concurrence (red dashed line) and classical correlations (green line)
for several values of the parameter $\gamma$.}
\label{classical}
\end{figure}

As mentioned above, classical correlations are much smaller than the concurrence for most of the evolution.  Depending on the analytical expression of the quantum discord ($D_{1}$ or $D_{2}$), we have two expressions for the classical correlations, each valid in a different time interval:
\begin{align}
CC_{1}(\tau)=&-2[a(\tau)+b(\tau)]\log_{2}[a(\tau)+b(\tau)]-\nonumber \\
&2[1-a(\tau)-b(\tau)]\log_{2}[1-a(\tau)-b(\tau)]+\nonumber \\
&[1-a(\tau)-2b(\tau)]\log_{2}[1-a(\tau)-2b(\tau)]+\nonumber \\
&a(\tau)\log_{2}a(\tau)+2b(\tau)\log_{2}b(\tau)
\end{align}
and,
\begin{align}
CC_{2}(\tau)=&-[a(\tau)+b(\tau)]\log_{2}[a(\tau)+b(\tau)]+ \nonumber \\
&a(\tau)\log_{2}a(\tau)+ [b(\tau)-c(\tau)]\log_{2}[b(\tau)-c(\tau)]+ \nonumber \\
&[b(\tau)+c(\tau)]\log_{2}[b(\tau)+c(\tau)]+ \nonumber \\
&[1-a(\tau)-2b(\tau)]\log_{2}[1-a(\tau)-2b(\tau)]+ \nonumber \\
&[a(\tau)+b(\tau)]\log_{2}[1-a(\tau)-b(\tau)]+ \nonumber \\
&\frac{1}{2}(1+\alpha)\log_{2}\left[\frac{1}{2}(1+\alpha)\right]\nonumber\\
&+ \frac{1}{2}(1-\alpha)\log_{2}\left[\frac{1}{2}(1-\alpha)\right]
\end{align}
where $a(\tau)$, $b(\tau)$, $c(\tau)$ are given in Eq.~(\ref{coeff}) and
$\alpha=[2(a(\tau)+b(\tau))-1]^2+4|c(\tau)|^2$.
Figure~\ref{classical} displays the classical correlations and the concurrence for several values of the parameter $\gamma$. Typically, classical correlations are of the order of 1\% of the concurrence.
\begin{acknowledgments}
The authors acknowledge financial support from the Brazilian funding agencies CNPq, CAPES,  FAPERJ, and also from the Brazilian National Institute of Science and Technology for Quantum Information.
\end{acknowledgments}

\end{document}